\documentclass[longauth]{aa} 
\usepackage{graphicx}
\usepackage{txfonts}
\usepackage{natbib}
\usepackage{color}
\usepackage[]{units}

\bibpunct{(}{)}{;}{a}{}{,} 

\newcommand{\hess}{\textsc{H.E.S.S.}}
\newcommand{\fermi}{\textsc{{\it Fermi}-LAT}}
\newcommand{\dgr}{\ensuremath{^\mathrm{o}}}
\newcommand{\numberObs}{47}

\newcommand{\hmsh}{\mbox{$^{\mathrm h}$}}%
\newcommand{\hmsm}{\mbox{$^{\mathrm m}$}}%
\newcommand{\hmss}{\mbox{$^{\mathrm s}$}}%
\newcommand{\HMS}[3]{$#1\hmsh#2\hmsm#3\hmss$}
\newcommand{\DMS}[3]{$#1\dgr#2\arcmin#3\arcsec$}

\begin{document} 

  \title{Flux upper limits for 47 AGN observed with \hess\ in 2004-2011}

	\author{H.E.S.S. Collaboration
	\and A.~Abramowski \inst{1}
	\and F.~Aharonian \inst{2,3,4}
	\and F.~Ait Benkhali \inst{2}
	\and A.G.~Akhperjanian \inst{5,4}
	\and E.~Ang\"uner \inst{6}
	\and G.~Anton \inst{7}
	\and S.~Balenderan \inst{8}
	\and A.~Balzer \inst{9,10}
	\and A.~Barnacka \inst{11}
% ...
	}

	\institute{
	Universit\"at Hamburg, Institut f\"ur Experimentalphysik, Luruper Chaussee 149, D 22761 Hamburg, Germany \and
	Max-Planck-Institut f\"ur Kernphysik, P.O. Box 103980, D 69029 Heidelberg, Germany \and
	Dublin Institute for Advanced Studies, 31 Fitzwilliam Place, Dublin 2, Ireland \and
	National Academy of Sciences of the Republic of Armenia, Yerevan  \and
	Yerevan Physics Institute, 2 Alikhanian Brothers St., 375036 Yerevan, Armenia \and
	Institut f\"ur Physik, Humboldt-Universit\"at zu Berlin, Newtonstr. 15, D 12489 Berlin, Germany \and
	Universit\"at Erlangen-N\"urnberg, Physikalisches Institut, Erwin-Rommel-Str. 1, D 91058 Erlangen, Germany \and
	University of Durham, Department of Physics, South Road, Durham DH1 3LE, U.K. \and
	DESY, D-15738 Zeuthen, Germany \and
	Institut f\"ur Physik und Astronomie, Universit\"at Potsdam,  Karl-Liebknecht-Strasse 24/25, D 14476 Potsdam, Germany \and
	Nicolaus Copernicus Astronomical Center, ul. Bartycka 18, 00-716 Warsaw, Poland
% ...
}
	
	\titlerunning{Flux upper limits for 47 AGN}
	\authorrunning{H.E.S.S. Collaboration}

   \date{}

  \abstract
   {About 40\% of the observation time of the High Energy Stereoscopic System (H.E.S.S.)  is dedicated to studying active galactic nuclei (AGN), with the aim of increasing the sample of known extragalactic very-high-energy (VHE, $E>100{\rm\ GeV}$) sources and constraining the physical processes at play in potential emitters.}
   {H.E.S.S. observations of AGN, spanning a period from April 2004 to December 2011, are investigated to constrain their $\gamma$-ray fluxes. Only the 47 sources without significant excess detected at the position of the targets are presented.}
   {Upper limits on VHE fluxes of the targets were computed and a search for variability was performed on the nightly time scale.}
   {For 41 objects, the flux upper limits we derived are the most constraining reported to date. These constraints at VHE are compared with the flux level expected from extrapolations of {\it Fermi}-LAT measurements in the two-year catalog of AGN. The H.E.S.S. upper limits are at least a factor of two lower than the extrapolated {\it Fermi}-LAT fluxes for 11 objects. Taking into account the attenuation by the extragalactic background light reduces the tension for all but two of them, suggesting intrinsic curvature in the high-energy spectra of these two AGN.}
   {Compilation efforts led by current VHE instruments are of critical importance for target-selection strategies before the advent of the  Cherenkov Telescope Array, CTA.}
\offprints{\\Jonathan Biteau - email: biteau(at)in2p3.fr\\
David Sanchez - email: david.sanchez(at)lapp.in2p3.fr}

   \maketitle

\section{Introduction} \label{Intro}

Since the discovery of their extragalactic origin fifty years ago \citep{1963Natur.197.1040S}, the class of astrophysical sources called active galactic nuclei (AGN) has been a prime target for astronomers that observe the sky from radio wavelengths to very-high-energy $\gamma$~rays (VHE, $E>100\ {\rm GeV}$). AGN are thought to host super-massive black holes (typical mass of $10^{8-9}M_\sun$) surrounded by an accretion disk, with a fraction of them showing two-sided jets. To unify the various subclasses of AGN, a scheme to sort them based on their orientation with respect to the observer's line of sight has been proposed since the 1990s \citep{Urry}. Objects whose jets are closely aligned with the line of sight are called blazars. They fall into two source classes, broad-line-spectrum sources called flat spectrum radio quasars (FSRQ), and BL Lac objects (hereafter BL in tables), which show faint lines or featureless spectra.

Active galactic nuclei, in particular blazars, are the most numerous objects detected at high energy (HE, $100\ {\rm MeV}<E<100\ {\rm GeV}$), where all-sky surveys can be performed with pair-conversion detectors onboard satellites such as the \textit{Fermi} Large Area Telescope \citep[LAT,][]{2009ApJ...697.1071A}. The Second LAT AGN Catalog, hereafter 2LAC, comprises 886 off-plane (i.e. above a Galactic latitude of 10\dgr) point-like sources associated with AGN that were detected in two years of operation \citep[][]{2LAC}. AGN constitute a third of the sources known at VHE, despite
a coverage biased toward Galactic sources. With the fast decrease of fluxes with increasing energy, observations at VHE are mostly performed with ground-based imaging atmospheric Cherenkov telescopes (IACT), which have a field of view (FoV) of a few degrees but an effective area on the order of $10^5\,{\rm m}^2$. Their current sensitivity prevents an all-sky scan in a reasonable amount of time, and IACT observations must be pointed to targets of interest (see e.g. \citealp{2013APh....43..317D} for a discussion of the capabilities of next generation instruments). Targeted AGN are selected based on their radio and X-ray spectra \citep{1996ApJ...473L..75S,2000AIPC..515...53P,2002AA...384...56C} as well as based on their HE flux extrapolated to VHE \citep{2010MNRAS.401.1570T}.

The High Energy Stereoscopic System \citep[\hess,][]{Crab} has significantly contributed to the expansion of the class of VHE AGN, with the detection of 23 objects, including 20 discoveries, among 56 known sources of this type as of the end of the year 2013\footnote{TeVCat, http://tevcat.uchicago.edu/}. The \hess\ experiment is located in the Khomas Highland, Namibia (23\dgr16'18'' S, 16\dgr30'01''E) at an altitude of $\unit[1800]{m}$ above sea level. In its first phase, this experiment was an array of four identical telescopes with cameras composed of 960 photomultipliers and segmented reflectors paving a reflective area of $\unit[107]{m^2}$, for an equivalent diameter of $\unit[12]{m}$. Most of the AGN detected with \hess\ belong to the BL Lac class, as shown in Table~\ref{Table:Detection}, with the exception of the two nearby radio galaxies of Fanaroff-Riley~I type (FR~I) Centaurus~A and M~87, the FSRQ PKS~1510-089, and the blazar candidate HESS J1943+213, which is located in the Galactic plane. In addition to constraining the radiative processes responsible for the $\gamma$-ray emission (for detailed discussions, see, e.g., the references in Table~\ref{Table:Detection}), the VHE spectra of these objects can also serve cosmological purposes, as shown with the constraints \citep{EBLAHA,2007A&A...471..439M} and indirect measurement \citep{2013A&A...550A...4H} of the extragalactic background light (EBL). With peak intensities in the optical and far-infrared bands, the EBL is composed of the integrated emission of stars and galaxies as well as of the reprocessing of UV-to-optical light by dust. The EBL can hardly be measured directly, although it is the second-most intense diffuse radiation in the Universe after the cosmic microwave background.

\begin{table*}
\centering
\small
\begin{tabular}{ l c c c }
\hline\hline
Object & $z$ & Type & Reference\\ 
\hline
Cen A & 0.002 & FR I & \cite{2009ApJ...695L..40A}\\ 
M 87 & 0.004 & FR I & \cite{2012ApJ...746..151A}\\ 
Markarian 421 & 0.031 & BL & \cite{MrkHESS2004}\\ 
AP Librae & 0.049 & BL & \cite{2012IAUS..284..411S} \\ 
PKS 1440-389 & 0.065 & BL & \cite{2012ATel.4072....1C} \\
PKS 0548-322 & 0.069 & BL & \cite{2010AA...521A..69A} \\ 
PKS 2005-489 & 0.071 & BL & \cite{2011AA...533A.110H} \\ 
RGB J0152+017 & 0.080 & BL & \cite{2008AA...481L.103A} \\ 
SHBL J001355.9-185406 & 0.095 & BL & \cite{SHBL} \\
1ES 1312-423 & 0.105 & BL & \cite{2013MNRAS.434.1889H} \\ 
PKS 2155-304 & 0.116 & BL & \cite{2012AA...539A.149H} \\ 
1ES 0229+200 & 0.140 & BL & \cite{0229}\\ 
1RXS J101015.9-311909 & 0.143 & BL & \cite{2012AA...542A..94H}\\
H 2356-309 & 0.165 & BL & \cite{2010AA...516A..56H}\\ 
1ES 1101-232 & 0.186 & BL & \cite{1101}\\ 
1ES 0347-121 & 0.188 & BL & \cite{0347}\\ 
PKS 0301-243 & 0.266 & BL & \cite{2013AA...559A.136H} \\ 
1ES 0414+009 & 0.287 & BL & \cite{2012AA...538A.103H}\\ 
PKS 1510-089 & 0.361 & FSRQ & \cite{1510} \\ 
PKS 0447-439 & $<0.57\ddagger$ & BL & \cite{2013AA...552A.118H}\\ 
PG 1553+113 & - & BL & \cite{2008AA...477..481A}\\ 
HESS J1943+213 & - & - & \cite{2011AA...529A..49H}\\ 
KUV 00311-1938 & $>0.506\dagger\ $ & BL & \cite{becherini:490} \\ 
\hline
\end{tabular}
\tablebib{$\dagger$ see \citet{2012AIPC.1505..566P}. $\ddagger$ see \citet{2013AA...552A.118H}.}
\caption{AGN detected by \hess\ as of September 2013. The redshift, classification and latest \hess\ publication on the source are given in columns two to four. Acronyms are defined in the text.}
\label{Table:Detection}
\end{table*}

During the eight years of the first phase of \hess, some of the observations did not result in significant excesses at the position of the target or in the FoV of the telescopes. A first set of upper limits \citep[][hereafter HUL1]{2005AA...441..465A} on 19 AGN observed during 63 hours was published after two years of observation. A second paper \citep[][hereafter HUL2]{2008AA...478..387A} listed 14 upper limits based on 94 hours of observation spanning 2005-2007. In this paper, which follows extensive compilation efforts from previous-generation instruments such as Whipple \citep{2004ApJ...603...51H} or HEGRA \citep{2004AA...421..529A}, 47 selected candidates are studied, with observations spanning April 2004 to December 2011, for a total live time of approximately 400 hours. The candidates and the data selection are presented in Sec.~\ref{Obs}, the event and spectral analyses are examined in Sec.~\ref{Ana}, and the constraints on the VHE emission are discussed in Sec.~\ref{Disc}, together with the target-selection strategy.

\section{Selected candidate VHE emitters} \label{Obs}

The sample of targets consists of the AGN observed with \hess\ until December 2011, for which more than an hour of corrected live time was recorded (see Sec.~\ref{Ana}). Only objects located away from the Galactic plane, that is above a Galactic latitude of 10\dgr, were taken into account. Neither datasets on potential or detected \hess\ sources\footnote{A source is considered as detected above a significance of $5\sigma$, while a potential source corresponds to an extrapolated observation time needed to reach detection shorter than $\unit[40]{hours}$. The list of objects studied in this paper does not depend on the latter criterion within $\pm\unit[10]{hours}$.} are included, nor those where upper limits based on the entire dataset have already been published (HUL1, HUL2). The objects listed in the 2LAC that are located in the same FoV as selected targets and are potentially associated with AGN were also studied. These criteria yield a total of 42 AGN and 5 unknown-type \fermi\ or EGRET sources, as listed in Table~\ref{Tab:ListOfTargets}. Pointed observations were performed for 33 objects, while 14 are visible in the FoVs. Of these 47 targets, 39 are studied for the first time with \hess\ in this paper, while eight of them (IIIZw 2, 1ES~0323+022, 3C~120, Pictor~A, 1ES~1440+122, RBS~1888, NGC~7469, 1ES~2343-151) have been re-observed since the publication of HUL1 and HUL2.

\begin{table*}
\centering
\small
\begin{tabular}{l c c c c c}
\hline\hline
Object & $\alpha_{\rm J2000}$ & $\delta_{\rm J2000}$ & $z$ & Type & Redshift reference \\ 
\hline
IIIZw 2 &  \HMS{00}{10}{31.2} & \DMS{+10}{58}{12} & 0.09 & FSRQ & \cite{2011MNRAS.414..500H} \\
1FGL J0022.2-1850 & \HMS{00}{22}{16.8} & \DMS{-18}{51}{00} & $>$0.77& BL   & \cite{2013ApJ...764..135S}\\
 & & & $<$1.38  &   & \cite{2012AA...538A..26R} \\
2FGL J0048.8-6347 & \HMS{00}{48}{52.8} & \DMS{-63}{48}{00} &  -  & - & - \\ 
PKS 0048-097 & \HMS{00}{50}{40.8} & \DMS{-09}{28}{48} & 0.64 & BL & \cite{2012AA...538A..26R}\\ 
1FGL J0051.4-6242 & \HMS{00}{51}{31.2} & \DMS{-62}{42}{36} & $<$1.12 & BL & \cite{2012AA...538A..26R} \\ 
RGB J0109+182 & \HMS{01}{09}{07.2} & \DMS{+18}{16}{12} & 0.14 & BL  & \cite{2000ApJS..129..547B} \\ 
2FGL J0211.2+1050 & \HMS{02}{11}{14.4} & \DMS{+10}{50}{24} & 0.20 & BL  & \cite{2010ApJ...712...14M} \\ 
2EG J0216+1107 & \HMS{02}{16}{00.0} & \DMS{+11}{07}{12} & - & -  & -  \\ 
2FGL J0229.3-3644 & \HMS{02}{29}{21.6} & \DMS{-36}{43}{48} & 2.12 & FSRQ  & \cite{2003AA...399..469H} \\ 
RBS 334  & \HMS{02}{37}{33.6} & \DMS{-36}{03}{36} & 0.41$\dagger$ & BL & \cite{2012AIPC.1505..566P} \\ 
RBS 0413 & \HMS{03}{19}{52.8} & \DMS{+18}{45}{36} & 0.19 & BL  &  \cite{2001AA...375..739D} \\
RBS 421 & \HMS{03}{25}{40.8} & \DMS{-16}{46}{12} & 0.29 & BL  &  \cite{2000ApJS..129..547B} \\ 
1ES 0323+022 & \HMS{03}{26}{14.4} & \DMS{+02}{25}{12} & 0.15 & BL & \cite{1999ApJ...525..127L} \\ 
QSO B0331-362  & \HMS{03}{33}{12.0} & \DMS{-36}{19}{48} & 0.31 & BL & \cite{2005ApJ...631..762W} \\ 
2FGL J0334.3-3728 & \HMS{03}{34}{19.2} & \DMS{-37}{28}{12} & $<$1.34 & BL  & \cite{2012AA...538A..26R} \\ 
PKS 0352-686 & \HMS{03}{52}{57.6} & \DMS{-68}{31}{12} & 0.09 & BL  & \cite{2011MNRAS.416.2840L} \\ 
2FGL J0426.6+0509c & \HMS{04}{26}{40.8} & \DMS{+05}{09}{00} & 1.33 & FSRQ & \cite{1999AAS..139..545K} \\ 
3C 120 & \HMS{04}{33}{12.0} & \DMS{+05}{21}{00} & 0.03 & FR I & \cite{2011MNRAS.416.2840L} \\ 
2FGL J0505.8-0411 & \HMS{05}{05}{48.0} & \DMS{-04}{12}{00} & 1.48 & FSRQ &  \cite{2001AJ....121.2843B} \\ 
1FGL J0506.9-5435 & \HMS{05}{06}{57.6} & \DMS{-54}{36}{00} & $<$1.07 & AGU & \cite{2012AA...538A..26R} \\ 
1ES 0507-040 & \HMS{05}{09}{38.4} & \DMS{-04}{00}{36} & 0.31 & BL & \cite{2005ApJ...631..762W} \\ 
2FGL J0515.0-4411 & \HMS{05}{15}{00.0} & \DMS{-44}{12}{00} & - &  - & - \\ 
2FGL J0516.5-4601 & \HMS{05}{16}{33.6} & \DMS{-46}{01}{12} & 0.19 & FSRQ & \cite{2004MNRAS.351...83L} \\ 
Pictor A & \HMS{05}{19}{50.4} & \DMS{-45}{46}{48} & 0.03 & FR II & \cite{2002AA...381..757L} \\ 
2FGL J0537.7-5716 & \HMS{05}{37}{43.2} & \DMS{-57}{16}{12} & 1.55 & AGU & \cite{2012AA...538A..26R} \\ 
2FGL J0540.4-5415 & \HMS{05}{40}{26.4} & \DMS{-54}{15}{00} & 1.19 & FSRQ & \cite{2008ApJS..175...97H} \\ 
BZB J0543-5532 & \HMS{05}{43}{57.6} & \DMS{-55}{31}{48} & 0.27 & BL & \cite{2012AIPC.1505..566P} \\ 
1ES 0715-259 & \HMS{07}{18}{04.8} & \DMS{-26}{08}{24} & 0.47 & BL  & \cite{2003AA...412..651C} \\ 
RBS 1049 & \HMS{11}{54}{04.8} & \DMS{-00}{10}{12} & 0.25 & BL & \cite{2008yCat.2282....0A} \\ 
1ES 1218+30.4 & \HMS{12}{21}{21.6} & \DMS{+30}{10}{48} & 0.18 & BL & \cite{2009yCat.2294....0A} \\ 
2FGL J1226.0+2953 & \HMS{12}{26}{04.8} & \DMS{+29}{54}{00} & - & - & - \\ 
3C 279 & \HMS{12}{56}{12.0} & \DMS{-05}{47}{24} & 0.54 & FSRQ & \cite{2006ApJ...638..642B} \\ 
1ES 1332-295 & \HMS{13}{35}{28.8} & \DMS{-29}{50}{24} & 0.26 & BL & \cite{2009MNRAS.399..683J} \\ 
PKS 1345+125 & \HMS{13}{47}{33.6} & \DMS{+12}{17}{24} & 0.12 & Sey II & \cite{2009yCat.2294....0A} \\ 
2FGL J1351.4+1115 & \HMS{13}{51}{28.8} & \DMS{+11}{15}{36} & 0.40 & BL & \cite{2009yCat.2294....0A} \\ 
1ES 1440+122 & \HMS{14}{42}{48.0} & \DMS{+12}{00}{36} & 0.16 & BL & \cite{2003AA...412..651C} \\ 
2FGL J1959.1-4245 & \HMS{19}{59}{09.6} & \DMS{-42}{45}{36} & 2.17 & FSRQ & \cite{2011MNRAS.411..901G} \\ 
PKS 2004-447 & \HMS{20}{07}{55.2} & \DMS{-44}{34}{48} & 0.24 & FSRQ & \cite{2009AA...495..691M} \\ 
RBS 1752  & \HMS{21}{31}{36.0} & \DMS{-09}{15}{36} & 0.45 & BL & \cite{2005AA...434..385G} \\ 
PG 2209+184 & \HMS{22}{11}{52.8} & \DMS{+18}{42}{00} & 0.07 & FSRQ & \cite{2002LEDA.........0P} \\ 
2FGL J2219.1+1805 & \HMS{22}{19}{12.0} & \DMS{+18}{05}{24} & 1.80 & FSRQ &  \cite{2005ApJ...626...95S} \\ 
RBS 1888 & \HMS{22}{43}{43.2} & \DMS{-12}{31}{12} & 0.23 & BL & \cite{1998AN....319..347F} \\ 
3EG J2248+1745 & \HMS{22}{48}{57.6} & \DMS{+17}{46}{12} & - & - & - \\ 
NGC 7469 & \HMS{23}{03}{16.8} & \DMS{+08}{52}{12} & 0.02 & Sey I & \cite{1999PASP..111..438F} \\ 
PMN J2345-1555 & \HMS{23}{45}{12.0} & \DMS{-15}{55}{12} & 0.62 & FSRQ & \cite{2008ApJS..175...97H} \\ 
1ES 2343-151 & \HMS{23}{45}{38.4} & \DMS{-14}{49}{12} & 0.22 & BL  & \cite{1993ApJ...412..541S} \\ 
2FGL J2347.9-1629 & \HMS{23}{47}{55.2} & \DMS{-16}{29}{24} & 0.58 & FSRQ & \cite{2002LEDA.........0P} \\ 
\hline
\end{tabular}
\tablebib{$\dagger$ Potential systematic uncertainties on the redshift of RBS~334 are discussed in \cite{2012AIPC.1505..566P}.}
\caption{Selected extragalactic objects observed with \hess\ from April 2004 to December 2011. Acronyms are defined in the text.}
\label{Tab:ListOfTargets}
\end{table*}

\begin{table*}
\centering
\small
\begin{tabular}{l r r r r r r r r r r}
\hline\hline
Object & T & Z$_{\rm obs}$ & Offset & T (corr.) & $E_{\rm th}$ & ON & OFF & $\alpha$ & Excess & S\\ 
 & [h] & \dgr & \dgr & [h] & [TeV] & & & & & [$\sigma$]\\
\hline
IIIZw 2 & 13.1 & 37 & 0.5 & 12.0 & 0.39 & 51 & 633 & 0.083 & -1.7 & -0.2\\ 
1FGL J0022.2-1850 & 61.5 & 13 & 2.1 & 15.4 & 0.24 & 104 & 6348 & 0.018 & -13.1 & -1.2\\ 
2FGL J0048.8-6347 & 8.0 & 40 & 1.2 & 4.9 & 0.58 & 23 & 431 & 0.033 & 8.8 & 2.1\\ 
PKS 0048-097 & 44.3 & 19 & 1.9 & 14.8 & 0.26 & 76 & 3418 & 0.023 & -3.2 & -0.4\\ 
1FGL J0051.4-6242 & 8.0 & 40 & 0.5 & 7.4 & 0.58 & 10 & 193 & 0.083 & -6.1 & -1.6\\ 
RGB J0109+182 & 4.1 & 42 & 0.5 & 3.8 & 0.71 & 10 & 144 & 0.083 & -2.0 & -0.6\\ 
2FGL J0211.2+1050 & 7.4 & 43 & 1.5 & 3.6 & 0.48 & 18 & 518 & 0.027 & 4.2 & 1.1\\ 
2EG J0216+1107 & 7.4 & 43 & 1.2 & 4.7 & 0.48 & 15 & 543 & 0.038 & -5.7 & -1.3\\ 
2FGL J0229.3-3644 & 6.1 & 14 & 1.8 & 2.0 & 0.39 & 7 & 421 & 0.021 & -1.8 & -0.6\\ 
RBS 334  & 6.1 & 14 & 0.5 & 5.6 & 0.35 & 26 & 293 & 0.083 & 1.6 & 0.3\\ 
RBS 0413 & 4.1 & 43 & 0.5 & 3.7 & 0.71 & 10 & 102 & 0.083 & 1.5 & 0.5\\ 
RBS 421 & 14.4 & 9 & 0.5 & 13.3 & 0.29 & 92 & 1153 & 0.083 & -4.1 & -0.4\\ 
1ES 0323+022 & 10.0 & 27 & 0.5 & 9.3 & 0.26 & 78 & 985 & 0.083 & -4.1 & -0.4\\ 
QSO B0331-362  & 30.6 & 19 & 1.1 & 20.6 & 0.24 & 109 & 3166 & 0.038 & -12.6 & -1.1\\ 
2FGL J0334.3-3728 & 24.7 & 18 & 1.6 & 11.4 & 0.26 & 84 & 2656 & 0.025 & 16.6 & 1.9\\ 
PKS 0352-686 & 15.0 & 47 & 0.5 & 14.2 & 0.71 & 36 & 423 & 0.083 & 0.8 & 0.1\\ 
2FGL J0426.6+0509c & 11.9 & 30 & 1.7 & 5.1 & 0.29 & 47 & 2137 & 0.023 & -1.8 & -0.3\\ 
3C 120 & 11.9 & 30 & 0.5 & 11.1 & 0.29 & 108 & 1008 & 0.083 & 24.0 & 2.4\\ 
2FGL J0505.8-0411 & 8.3 & 21 & 1.1 & 5.8 & 0.29 & 54 & 1306 & 0.035 & 7.9 & 1.1\\ 
1FGL J0506.9-5435 & 2.1 & 32 & 0.5 & 2.0 & 0.95 & 2 & 41 & 0.083 & -1.4 & -0.8\\ 
1ES 0507-040 & 8.3 & 21 & 0.5 & 7.7 & 0.32 & 52 & 614 & 0.083 & 0.8 & 0.1\\ 
2FGL J0515.0-4411 & 20.9 & 29 & 1.8 & 7.4 & 0.24 & 61 & 2877 & 0.021 & 0.1 & 0.0\\ 
2FGL J0516.5-4601 & 20.9 & 29 & 0.8 & 17.1 & 0.26 & 132 & 2123 & 0.056 & 12.8 & 1.1\\ 
Pictor A & 20.9 & 29 & 0.5 & 19.4 & 0.29 & 134 & 1367 & 0.083 & 20.1 & 1.8\\ 
2FGL J0537.7-5716 & 8.8 & 33 & 2.0 & 2.7 & 0.35 & 19 & 1103 & 0.019 & -1.8 & -0.4\\ 
2FGL J0540.4-5415 & 8.8 & 33 & 1.5 & 4.7 & 0.35 & 26 & 1303 & 0.027 & -8.9 & -1.6\\ 
BZB J0543-5532 & 8.8 & 33 & 0.5 & 8.1 & 0.39 & 49 & 652 & 0.083 & -5.3 & -0.7\\ 
1ES 0715-259 & 5.7 & 13 & 1.9 & 1.9 & 0.32 & 15 & 788 & 0.021 & -1.4 & -0.4\\ 
RBS 1049 & 4.3 & 30 & 0.5 & 3.9 & 0.39 & 17 & 253 & 0.083 & -4.1 & -0.9\\ 
1ES 1218+30.4 & 2.3 & 56 & 0.5 & 2.1 & 1.41 & 12 & 85 & 0.083 & 4.9 & 1.6\\ 
2FGL J1226.0+2953 & 2.3 & 56 & 1.2 & 1.4 & 1.41 & 10 & 147 & 0.031 & 5.4 & 2.1\\ 
3C 279 & 5.5 & 26 & 0.5 & 5.0 & 0.29 & 35 & 475 & 0.075 & -0.5 & -0.1\\ 
1ES 1332-295 & 10.1 & 25 & 0.7 & 8.4 & 0.26 & 54 & 1059 & 0.054 & -2.9 & -0.4\\ 
PKS 1345+125 & 7.9 & 37 & 0.7 & 6.7 & 0.53 & 22 & 351 & 0.056 & 2.5 & 0.5\\ 
2FGL J1351.4+1115 & 7.9 & 37 & 1.6 & 3.6 & 0.48 & 7 & 531 & 0.026 & -6.6 & -2.0\\ 
1ES 1440+122 & 11.2 & 37 & 0.5 & 10.4 & 0.29 & 66 & 650 & 0.083 & 11.8 & 1.5\\ 
2FGL J1959.1-4245 & 12.9 & 33 & 2.1 & 2.7 & 0.39 & 8 & 994 & 0.016 & -8.1 & -2.2\\ 
PKS 2004-447 & 25.6 & 33 & 0.5 & 23.5 & 0.39 & 110 & 1139 & 0.083 & 15.1 & 1.4\\ 
 RBS 1752  & 25.1 & 16 & 0.5 & 23.1 & 0.29 & 149 & 2023 & 0.083 & -19.6 & -1.5\\ 
PG 2209+184 & 8.8 & 42 & 0.5 & 8.1 & 0.64 & 19 & 286 & 0.083 & -4.8 & -1.0\\ 
2FGL J2219.1+1805 & 8.8 & 42 & 1.9 & 2.6 & 0.64 & 7 & 529 & 0.019 & -3.2 & -1.1\\ 
RBS 1888 & 7.9 & 14 & 0.5 & 7.3 & 0.22 & 74 & 916 & 0.077 & 3.5 & 0.4\\ 
3EG J2248+1745 & 17.3 & 43 & 1.8 & 5.8 & 0.48 & 36 & 1069 & 0.024 & 10.0 & 1.8\\ 
NGC 7469 & 7.9 & 33 & 0.5 & 7.4 & 0.32 & 79 & 772 & 0.083 & 14.7 & 1.7\\ 
PMN J2345-1555 & 21.0 & 15 & 1.0 & 15.9 & 0.22 & 147 & 3775 & 0.037 & 6.4 & 0.5\\ 
1ES 2343-151 & 21.0 & 15 & 0.7 & 18.4 & 0.22 & 156 & 2629 & 0.066 & -18.5 & -1.4\\ 
2FGL J2347.9-1629 & 21.0 & 15 & 1.6 & 9.6 & 0.20 & 104 & 3593 & 0.025 & 15.1 & 1.5\\ 
\hline
\end{tabular}
\caption{Results from \hess\ observations of \numberObs\ AGN. The first five columns give the characteristics of the observation (target name, duration, zenith angle, average {\it wobble} offset and acceptance-corrected time). Column 6 is the energy threshold. The number of ON and OFF events above the energy threshold, and the normalization of the $\rm OFF$ events, $\alpha$, are shown in columns 7, 8 and 9. The resulting excess and significance are given in the last two columns.}
\label{Table:Analysis}
\end{table*}

\begin{table*}
\centering
\small
\begin{tabular}{l c c l c l}
\hline\hline
Object & $E_{th}$ & I($>E_{th}$) & I($>E_{th}$) & ${\cal P}(\chi^2)$ & MJD-50000\\
 & [TeV] & [$\times \unit[10^{-12}]{cm^{-2}\ s^{-1}}$] & [\% C.U.] & [\%] & \\
\hline
IIIZw 2 & 0.39 & 0.67 & 0.7$^{U}$ & 22 & 3943-44,3953,4267,4270,4272,4274-76,4279,4320, \\
& & & & & 4322-26, 4328,4331-33\\
1FGL J0022.2-1850& 0.24 & 0.85 & 0.4$^{U}$ & 79 & 4653-60,5064,5090-92,5094,5112,5115-18,5415,\\
& & & & & 5417,5419,5422-27,5443-44,5448-51,5482,5501,\\
& & & & & 5504,5506-08,5775-76,5783,5885,5887-91,5910,5912 \\
2FGL J0048.8-6347 & 0.58 & 1.18 & 2.3$^{U}$ & 50 & 5833-37\\
PKS 0048-097 & 0.26 & 0.88 & 0.5$^{U}$ & 69 & 4023,4050-57,4321-26,4328,4331-35,4349,4350, \\
 & & & & & 4352-53,4357,4359-60,4363,4374,4378-79,4381-85,  \\
 & & & & & 5058,5060,5063-65,5067-68\\
1FGL J0051.4-6242 & 0.58 & 0.47 & 0.9$^{U}$ & 32 & 5833-37\\
RGB J0109+182 & 0.71 & 0.67 & 1.8$^{U}$ & 15 & 5093,5095\\
2FGL J0211.2+1050 & 0.48 & 1.21 & 1.7$^{U}$ & 39 & 3966-69,3971-72,3974,3976-78\\
2EG J0216+1107 & 0.48 & 0.66 & 0.9$^{U}$ & 63 & 3966-67,3969,3971-72,3974,3976-78\\
2FGL J0229.3-3644 & 0.39 & 1.05 & 1.1$^{U}$ & 67 & 5444,5446,5448-52\\
RBS 334 & 0.35 & 1.47 & 1.3$^{U}$ & 31 & 5444,5446,5448-52\\
RBS 0413 & 0.71 & 0.80 & 2.2$^{D}$ & 22 & 5446,5448-51,5482-83\\
RBS 421 & 0.29 & 0.89 & 0.6$^{U}$ & 98 & 4715,4717,4720,4815,4818-20,4822-30\\
1ES 0323+022 & 0.26 & 1.29 & 0.7$^{U}$ & 84 & 3267-68,3675-77,3996-4000\\
QSO B0331-362 & 0.24 & 0.82 & 0.4$^{U}$ & 89 & 3590,3592,3594-95,3597-98,3623,3625-27,\\
& & & & & 3638-39, 3641-42,3643-44,4353,4358,4360-61, \\
& & & & & 4363-64,4379-86,4391\\
2FGL J0334.3-3728 & 0.26 & 1.77 & 0.9$^{U}$ & 14 & 3589-90,3592,3597-98,3623,3625-27,3637-38,3641-44, \\
 & & & & & 4353,4358,4360-61,4363-64,4378-86,4391\\
PKS 0352-686 & 0.71 & 0.40 & 1.1$^{U}$ & 43 & 5483-84,5499-5502,5504-08,5510-12,5526-27,5529, \\
 & & & & & 5532-37\\
2FGL J0426.6+0509c & 0.29 & 1.57 & 1.0$^{U}$ & 30 & 3315-17,3352-54,5834-39,5841-43,5867-68\\
3C 120 & 0.29 & 2.23 & 1.4& 73 & 3315-18,3352-54,5834-43,5867-68\\
2FGL J0505.8-0411 & 0.29 & 2.14 & 1.3$^{U}$ & 45 & 4439,4441-46,4450\\
1FGL J0506.9-5435 & 0.95 & 0.52 & 2.3$^{U}$ & 87 & 5867-68\\
1ES 0507-040 & 0.32 & 1.37 & 1.0$^{U}$ & 69 & 4439,4441-46,4450\\
2FGL J0515.0-4411 & 0.24 & 1.94 & 0.9$^{U}$ & 27 & 3268-70,3273,3318-19,3350,3352-53,4050-53, \\
 & & & & & 4055-56,4059-62,4496,4498-99,4819-20,4823\\
2FGL J0516.5-4601 & 0.26 & 1.63 & 0.9$^{U}$ & 89 & 3268-70,3273,3318-19,3350,3352-53,4051-53,\\
 & & & & & 4055-56,4059-62,4496,4499,4819-20,4823\\
Pictor A & 0.29 & 1.44 & 0.9$^{U}$ & 12 & 3268-70,3273,3318-19,3350,3352-53,4050-53,\\
 & & & & & 4055-56,4059-62,4496,4498-99,4819-20,4823\\
2FGL J0537.7-5716 & 0.35 & 2.03 & 1.7$^{U}$ & 78 & 5911,5914,5917,5922-25\\
2FGL J0540.4-5415 & 0.35 & 1.16 & 1.0$^{U}$ & 27 & 5911,5914,5917,5922-25\\
BZB J0543-5532 & 0.39 & 0.90 & 0.9$^{U}$ & 25 & 5911,5914,5917,5922-25\\
1ES 0715-259 & 0.32 & 2.09 & 1.5$^{U}$ & 96 & 4140-44,4146,4148\\
RBS 1049 & 0.39 & 1.16 & 1.2$^{U}$ & 22 & 5320-23\\
1ES 1218+30.4 & 1.41 & 0.97 & 8.0$^{D}$ & 19 & 3875-76\\
2FGL J1226.0+2953 & 1.41 & 1.31 & 11$^{U}$ & 90 & 3875-76\\
3C 279 & 0.29 & 1.85 & 1.2$^{D}$ & 40 & 4118-21,4501,4855,4858-59,4861\\
1ES 1332-295 & 0.26 & 1.53 & 0.8$^{U}$ & 45 & 3929-35\\
PKS 1345+125 & 0.53 & 0.68 & 1.1$^{U}$ & 21 & 4938-41,4944-46,4948-49,4952\\
2FGL J1351.4+1115 & 0.48 & 0.51 & 0.7$^{U}$ & 40 & 4938-41,4944-46,4948-49,4952\\
1ES 1440+122 & 0.29 & 1.66 & 1.0$^{D}$ & 47 & 3109,3119,4995-99,5002-03,5005-06\\
2FGL J1959.1-4245 & 0.39 & 1.01 & 1.0$^{U}$ & 95 & 5358-59,5362,5365,5367,5369,5386,5389-91,5393-94, \\
 & & & & & 5396-97,5413,5415-16,5419,5421-23\\
PKS 2004-447 & 0.39 & 0.88 & 0.9$^{U}$ & 26 & 5358-59,5361-62,5364-67,5369-70,5386-87,5389-90, \\
 & & & & & 5391-96,5413-16,5418-24\\
RBS 1752 & 0.29 & 0.56 & 0.3$^{U}$ & 37 & 4625-32,4653-56,4728-39\\
PG 2209+184 & 0.64 & 0.38 & 0.9$^{U}$ & 52 & 4373,4375-76,4378-79,4381-86\\
2FGL J2219.1+1805 & 0.64 & 0.42 & 1.0$^{U}$ & 47 & 4374,4376-79,4381-86\\
RBS 1888 & 0.22 & 2.16 & 0.9$^{U}$ & 94 & 3207-10,3914-18\\
3EG J2248+1745 & 0.48 & 1.10 & 1.6$^{U}$ & 99 & 4292-96,4298-04,5004-09\\
NGC 7469 & 0.32 & 1.80 & 1.3 & 70 & 3202,3206,3211-12,4019-20,4022-23\\
PMN J2345-1555 & 0.22 & 1.65 & 0.7$^{U}$ & 47 & 3211-13,3590,3592-95,3597-98,5495-96,5498-99\\
1ES 2343-151 & 0.22 & 0.97 & 0.4$^{U}$ & 20 & 3212-13,3590,3592-93,3594-95,3597-98,5495-96, \\
 & & & & & 5498-99\\
2FGL J2347.9-1629 & 0.20 & 3.15 & 1.1$^{U}$ & 88 & 3211-12,3590,3592-93,3594-95,3597-98,5495-96, \\
 & & & & & 5498-99\\
\hline
\end{tabular}
\caption{Spectral and temporal analysis of \numberObs\ AGN. The upper limits given in column 3 and 4 are computed at the 99.9\% level. The superscript $^U$ indicates the best VHE upper limit computed for this target to date and $^D$ corresponds to a source detected by another VHE instrument. The observation nights are listed in the last column, and the $\chi^2$ probabilities for a constant fit of the flux at this time scale are shown in column 5.}
\label{Table:UL}
\end{table*}

\begin{figure*}[hdtp]
\includegraphics[width=0.45\linewidth]{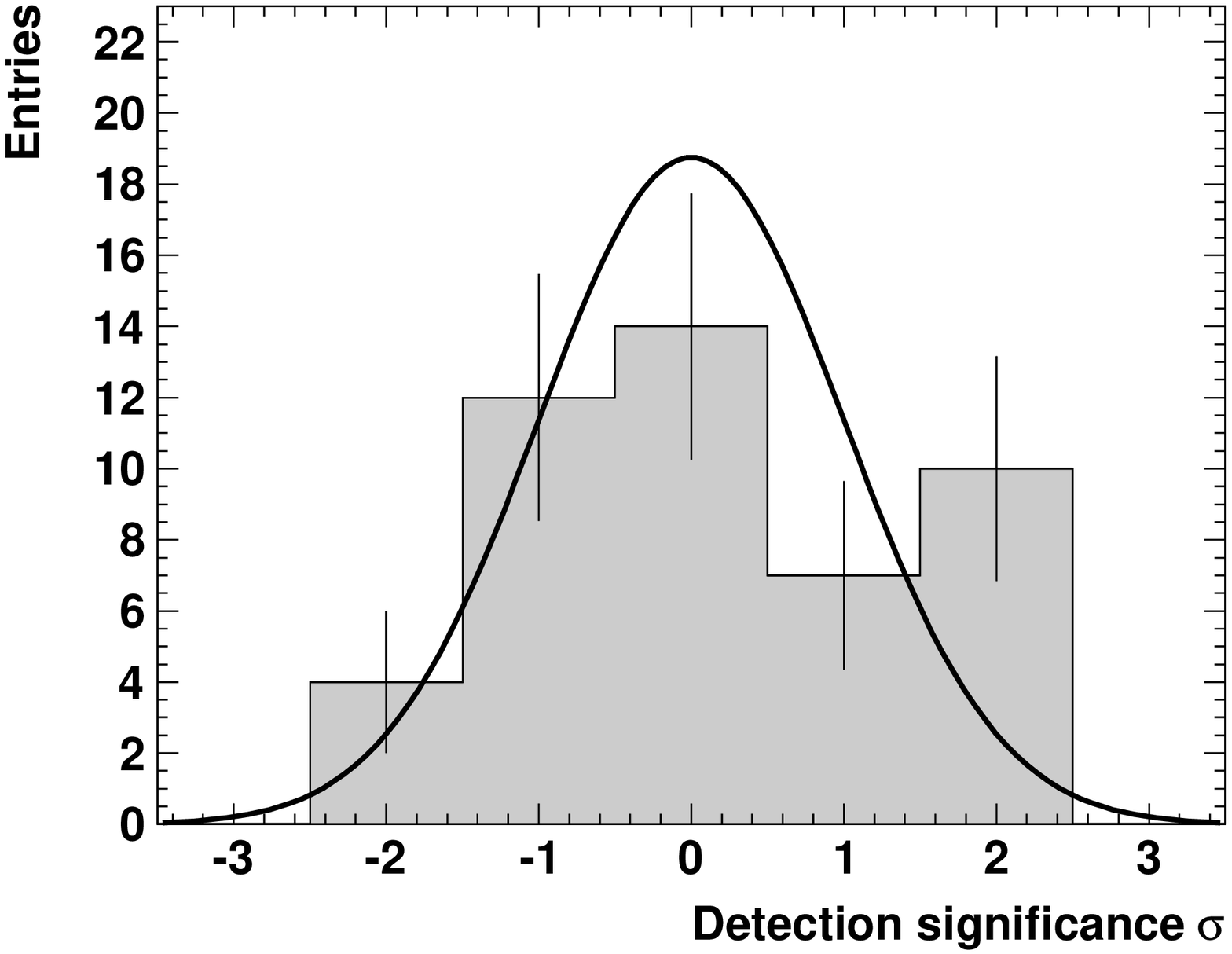} \hfill \includegraphics[width=0.45\linewidth]{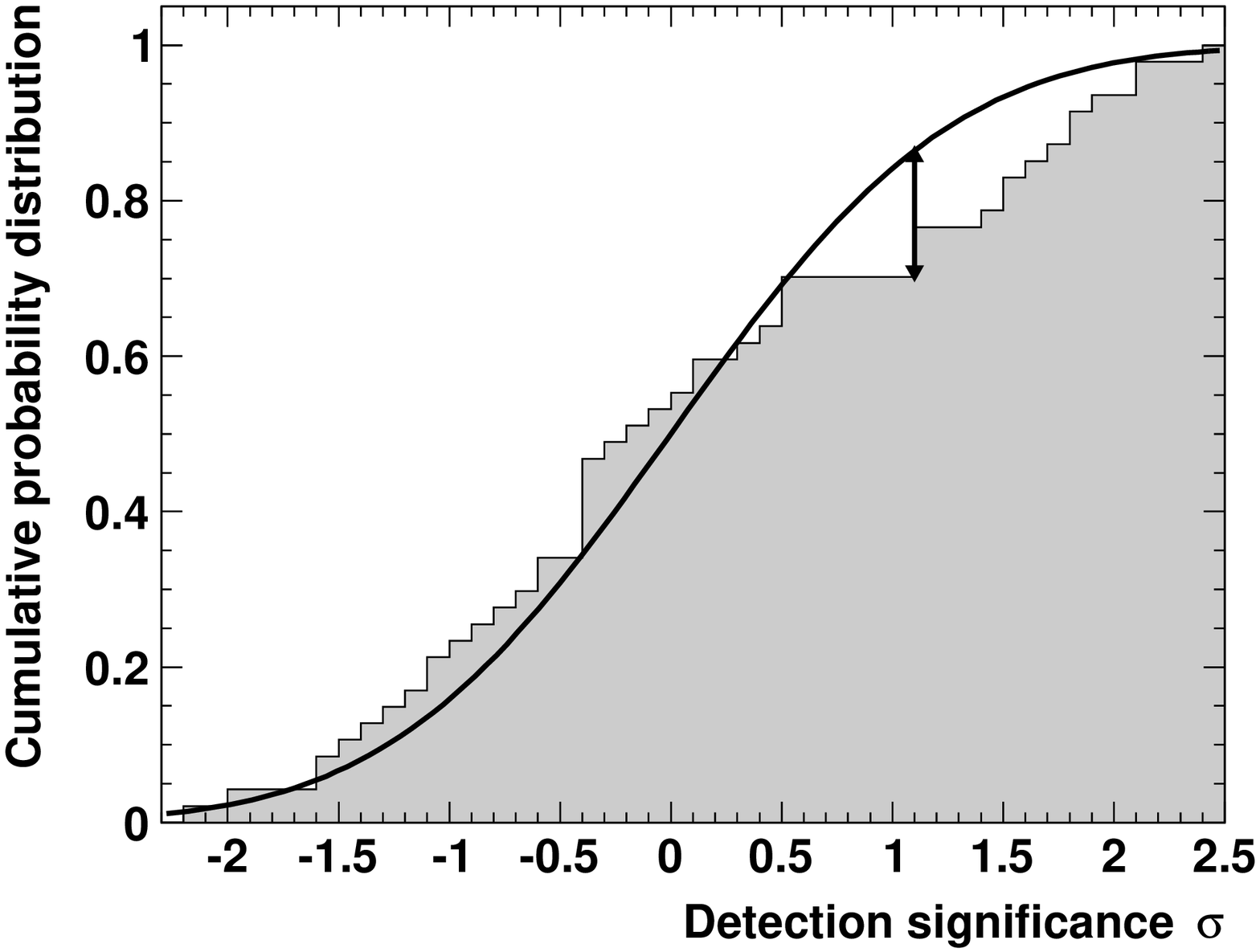}
\caption{{\it Left:} Distribution of the detection significances, $\sigma$, for the \numberObs\ candidates using $1\sigma$-wide bins. Error bars indicate the square root of the number of events in each bin, and the black line is a normal distribution of 47 events centered on zero and of unit width. {\it Right:} Cumulative distribution function of the detection significance and of the normal distribution. The maximum distance between the distributions is shown as a double-sided arrow.}
\label{FIG:Sig_Dis}
\end{figure*}

The redshifts of the targets were extracted from the Roma-BZCAT catalog Ed. 4.1.1 \citep{2009AA...495..691M}, from the work of \cite{2012AA...538A..26R} on AGN detected with \fermi, and from the publication by \cite{2012AIPC.1505..566P} about VHE (candidate) emitters. The redshifts of objects not listed in these publications were individually searched for in the literature. A detailed list of references can be found in the last column of Table~\ref{Tab:ListOfTargets}. The distant objects with $z>1$ in this table were not directly targeted. 2FGL~J0426.6+0509c is located in the same FoV as 3C~120. 2FGL~J1959.1-4245 and 2FGL~J2219.1+1805 are neighbors of PKS~2004-447 and RBS~1888, respectively.  2FGL~J0505.8-0411 and 2FGL~J0540.4-5415 were jointly observed with BZB~J0543-5532.

The classification of the targets is also primarily based on BZCAT. For objects not listed there, the 2LAC catalog was followed, yielding two AGN of unknown type in addition to BL Lac and FSRQ objects: 1FGL~J0506.9-5435 and 2FGL~J0537.7-5716, called AGU in Table~\ref{Tab:ListOfTargets}, following the 2LAC nomenclature. Following HUL1 and HUL2, 3C~120 and Pictor~A are classified as Fanaroff-Riley~I (FR~I) and II (FR~II) radio galaxies. Searching the SIMBAD database\footnote{http://simbad.u-strasbg.fr/simbad/}, Seyfert~1 nuclei (Sey I) are hosted by these two objects and by NGC~7469, while PKS~1345+125 is classified as a Seyfert~2 (Sey II). To summarize, most of the targets are blazars, with 13 FSRQ and 23 BL Lac objects, including PKS~0352-686 and 1FGL~J0022.2-1850 recently confirmed as BL Lac objects by \cite{2009AA...494..417R} and \cite{2013ApJ...764..135S}, respectively.

\begin{table*}
\centering
\small
\begin{tabular}{ l c c c c c }
\hline\hline
Object & $z$ & $E_{th}$ & I($ >E_{th}$) & I$_{2LAC}(>E_{th})$ &  I$^{\rm EBL}_{2LAC}(>E_{th})$ \\ 
 & & [TeV] & [\% C.U.] & [\% C.U.] & [\% C.U.] \\
\hline
2FGL J1351.4+1115 & 0.40 & 0.48 & 0.7 & 40 & 0.2\\
1FGL J0022.2-1850 & 0.77-1.38 & 0.24 & 0.4 & 24 & 0.1\\
1FGL J0051.4-6242 & $<$1.12 & 0.58 & 0.9 & 37 & 0.5\\
BZB J0543-5532 & 0.27 & 0.39 & 0.9 & 25 & 1.4\\
1FGL J0506.9-5435 & $<$1.07 & 0.95 & 2.3 & 65 & 0.2\\
RBS 334  & 0.41 & 0.35 & 1.3 & 13 & 0.2 \\
PKS 0048-097 & 0.64 & 0.26 & 0.5 & 3.6 & 0.05\\
2FGL J0334.3-3728 & $<$1.34 & 0.26 & 0.9 & 7.3 & 1.0\\
RBS 1049 & 0.25 & 0.39 & 1.2 & 5.0 & 0.4\\
PMN J2345-1555 & 0.62 & 0.22 & 0.7 & 2.7 & 0.1\\
RBS 421 & 0.29 & 0.29 & 0.6 & 1.8 & 0.2\\
RBS 1752  & 0.45 & 0.29 & 0.3 & 1.0 & 0.04\\
\hline
\end{tabular}
\caption{Comparison of the high-energy extrapolation from the 2LAC with \hess\ upper limits. Only objects with constraining limits are selected. I$_{2LAC}(>E_{th})$ and I$^{\rm EBL}_{2LAC}(>E_{th})$ are the 2LAC measurements extrapolated above $E_{th}$, taking into account the EBL absorption for the second quantity. When only an upper limit on the redshift is available, a value of $z=0.3$ is assumed to derive these extrapolations. For 1FGL~J0022.2-1850, the lower limit $z>0.77$ is used. }
\label{Table:Extrapol}
\end{table*}

\section{Analysis and results} \label{Ana}

The observation conditions and the results of the event analyses are listed in Table~\ref{Table:Analysis}.
The \hess\ telescopes are usually pointed with an offset angle of $0.5-0.7\dgr$ (\textit{wobble} mode) when observing extragalactic sources. Higher offset values occur in Table~\ref{Table:Analysis} because sources can be in the same FoV as a scheduled target source. The comparably (with other IACTs) large FoV of 5\dgr\ of \hess\ telescopes allows for reliable spectral reconstruction up to an offset of at least 2\dgr\ \citep[offset value in][]{2013MNRAS.434.1889H}, close to the maximum offset values that are listed in Table~\ref{Table:Analysis}. The observation time, shown in the second column, is corrected for the decrease in acceptance due to an increasing offset from the centre of the cameras. This correction results in a shorter acceptance-corrected live time, as shown in column 5. 

The data that pass standard quality criteria \citep[good weather, stability of the instrument, as in][]{Crab} were analyzed with {\it Model++ Standard Cuts} \citep{MathieuANA}, corresponding to a selection criterion on the image charge of 60 photo electrons. A cross-check was performed with a multivariate analysis described in \citet{TMVAHD}. The results of the analysis of the 47 targets described in the following were derived with a single pipeline, associated to the {\it Model ++} analysis. The analysis energy threshold\footnote{Hereafter, the energy threshold is defined as the energy for which the acceptance reaches 20\% of the highest value. This approach, which results in a somewhat lower threshold than the conventional definition (peak of the energy distribution of the events), corresponds to an energy bias lower than the energy resolution \citep[see Fig.~23-24 in][]{MathieuANA}, which ensures the quality of the reconstructed spectrum.}, shown in column 6, depends on the average zenith angle of the observations (column 3) and on the offset from the center of the cameras (column 4). The number of ON-target (column 7) and OFF-target events (column 8) was measured above the threshold energy in regions of 0.1\dgr\ radius. The normalization $\alpha$ of the $\rm OFF$ events, shown in column 9, is a relative exposure normalization factor between the ON and OFF regions, within the {\it Reflected} background modeling method \citep{Crab,2007A&A...466.1219B}. The excess, defined as ${\rm ON}-\alpha\times{\rm OFF}$, and its significance, calculated using equation 17 in \cite{LiMa}, are shown in the last two columns of Table~\ref{Table:Analysis}. No significant deviation from zero is observed, with values in the range $[-2.2\sigma;2.4\sigma]$.

The distribution of the detection significance is compared in Fig.~\ref{FIG:Sig_Dis} with a normal distribution of 47 events, centered on zero and of unitary standard deviation. The deviations of the data distribution from the normal distribution were quantified using a Kolmogorov-Smirnov test. The highest value of the absolute difference between the cumulative probability distributions reaches 0.17, with a $p$-value for a normal distribution of 12\%, equivalent to a 1.5 Gaussian standard deviation. An Anderson-Darling test yields a similar result, with a $p$-value for a normal distribution of 10\%. These tests do not indicate a collective excess of events within the sample of source candidates.

As in HUL1 and HUL2, the spectral analysis was performed assuming a power-law spectrum with photon index $\Gamma = 3$, close to values observed for the sources listed for instance in Table~\ref{Table:Detection}. Upper limits on the integral fluxes above the threshold energies were computed at the 99.9\% confidence level, according to the statistical method of \cite{2005NIMPA.551..493R}. The limits shown in column 4 of Table~\ref{Table:UL} were converted into Crab units (C.U., column 5) using the power-law spectrum measured by \cite{Crab}, with a photon index $\Gamma = 2.63$ and flux at $\unit[1]{TeV}$ $\phi_0 = \unit[3.45\times10^{-11}]{cm^{-2}\ s^{-1}\ TeV^{-1}}$.

A search for variability, one of the characteristic properties of AGN, was performed by fitting a constant function to the flux estimates derived on a night-by-night time scale, as in HUL1 and HUL2. The modified Julian dates of observation for which at least one ON-event is recorded are given in the last column of Table~\ref{Table:UL} and the $\chi^2$ probabilities for a constant fit (with $N_{\rm nights}-1$ degrees of freedom) are shown in column 5. With $\chi^2$ probabilities higher than 10\%, no flaring event is detected on the nightly time scale. A search on shorter time scales is ruled out by the small statistics in each temporal bin, and larger bins would result, for some of the targets, in a number of points that is too small to lead such a study.

\section{Discussion} \label{Disc}

Among the 47 candidates, four blazars have been detected by other IACTs. The BL Lac object RBS~0413 has been discovered by the VERITAS Collaboration \citep{2012ApJ...750...94A} at the 1\% C.U. level, in agreement with the upper limit of 2.2\% C.U. set in this study. 1ES~1218+30.4, detected by the MAGIC \citep{2006ApJ...642L.119A} and VERITAS \citep{2009ApJ...695.1370A,2010ApJ...709L.163A} Collaborations, is a known variable BL Lac object, with reported fluxes between 6\% and 20\% C.U. These are on the order of the upper limit of 8.0\% measured above the comparably high energy threshold of $\unit[1.4]{TeV}$. The VHE flux of the FSRQ 3C~279 has been measured at the 0.5\% C.U. level by the MAGIC Collaboration \citep{2011A&A...530A...4A} and is compatible with the 1.2\% C.U. derived here. The last detected BL Lac object in the list of targets is 1ES~1440+122, with a flux of 1\% C.U. \citep{2011arXiv1110.0040W} that matches the upper limit derived in this paper.

The upper limits on 3C 120 and NGC~7469 are a factor of two higher than those derived in HUL1, despite a doubled amount of data. This can be related to background fluctuations, with negative detection significances of $\sim-2\sigma$ in HUL1 and $\sim+2\sigma$ upward fluctuations observed in this study.

For the other targets, that is 41 among the 47 AGN, the upper limits derived in Table~\ref{Table:UL} are the strongest reported to date\footnote{Variations in the energy thresholds of different instruments that observed the same targets were taken into account when comparing upper limits. Values are also reported in Crab units in this paper for the sake of clarity.}, with values down to 0.4\% C.U. These upper limits are compared with the HE flux reported in the 2LAC, extrapolated above the threshold energy of \hess, I$_{2LAC}(>E_{th})$, without taking into account absorption by the EBL. Since \hess\ observed the sky for a longer period than \fermi, the 2LAC spectra are not strictly simultaneous with the VHE constraints. The comparison of the \fermi\ extrapolated fluxes and of the \hess\ upper limits is thus based on the assumption that the 2LAC spectra are representative of the average HE emission. This assumption is corroborated by 2LAC studies of FSRQs and BL Lac objects that show an average fractional variance of the flux (square root of the normalized excess variance) on the order of $0.55\pm0.10$, that is fluxes that vary on average within $\pm 55\%$, and also by a rather short duty cycle for high flux events (above 1.5 standard deviation), with a most probable value for the duty cycle on the order of 5\% to 10\%. The targets for which I$_{2LAC}(>E_{th})$ is at least twice as high\footnote{A fiducial value of two corresponds to the average EBL absorption between $\unit[500]{GeV}$ and $\unit[1]{TeV}$ for a source situated at $z\sim0.1$ \citep[e.g., within the modeling of][]{Fran08}.} as the \hess\ upper limit are listed in Table~\ref{Table:Extrapol}. Sources detected with other IACTs as well as the distant 2FGL~J0537.7-5716 ($z=1.55$) are not included in the list.

The extrapolated fluxes of these sources are higher than the \hess\ upper limits, indicating curvature in their spectra. The curvature can have an intrinsic (i.e. related to the underlying emitting particles) and extrinsic (i.e. due to the EBL absorption) origin. To constrain the origin of this curvature, the \fermi\ fluxes, $\phi_{2LAC}(E)$, were extrapolated taking into account the best-fit EBL model derived with the \hess\ data, corresponding to the optical depth of \cite{Fran08}, $\tau_{\rm FR08}(E,z)$ scaled up by a factor $\alpha_0=1.27$ \citep{2013A&A...550A...4H}. The EBL-absorbed extrapolations are thus computed as I$^{\rm EBL}_{2LAC}(>E_{th}) = \int_{E_{th}}{\rm d}E\ \phi_{2LAC}(E)\ e^{-\alpha_0 \times \tau_{\rm FR08}(E,z)}$. Targets for which only an upper limit on the redshift was available were assumed to lie at $z=0.3$, roughly corresponding to the peak of the distribution for BL Lac objects in the 2LAC. For 1FGL~J0022.2-1850, the EBL-absorbed extrapolation derived using a redshift of $0.77$ does not exceed the \hess\ upper limit.

\begin{figure*}[hdtp]
\includegraphics[width=0.45\linewidth]{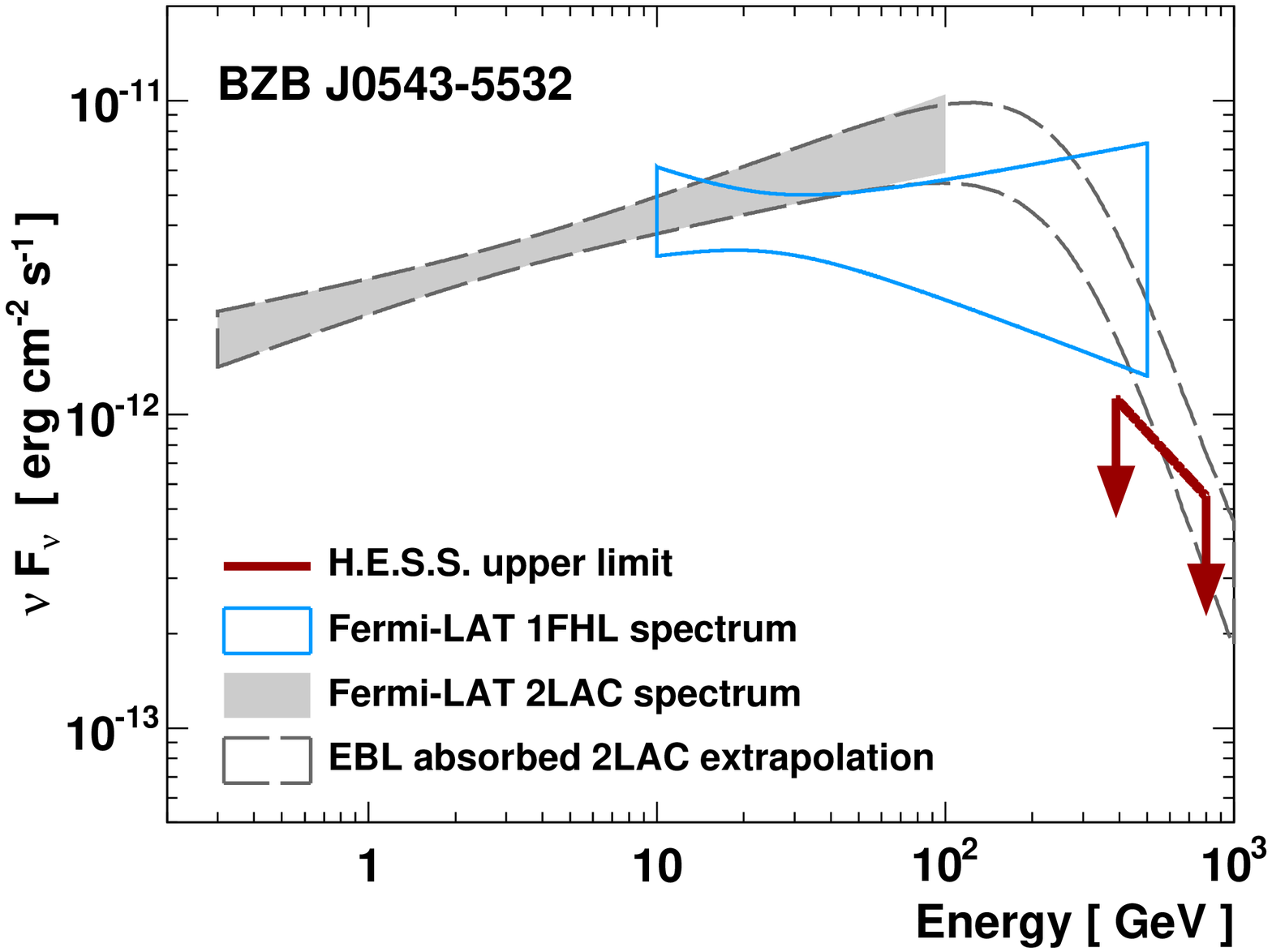} \hfill \includegraphics[width=0.45\linewidth]{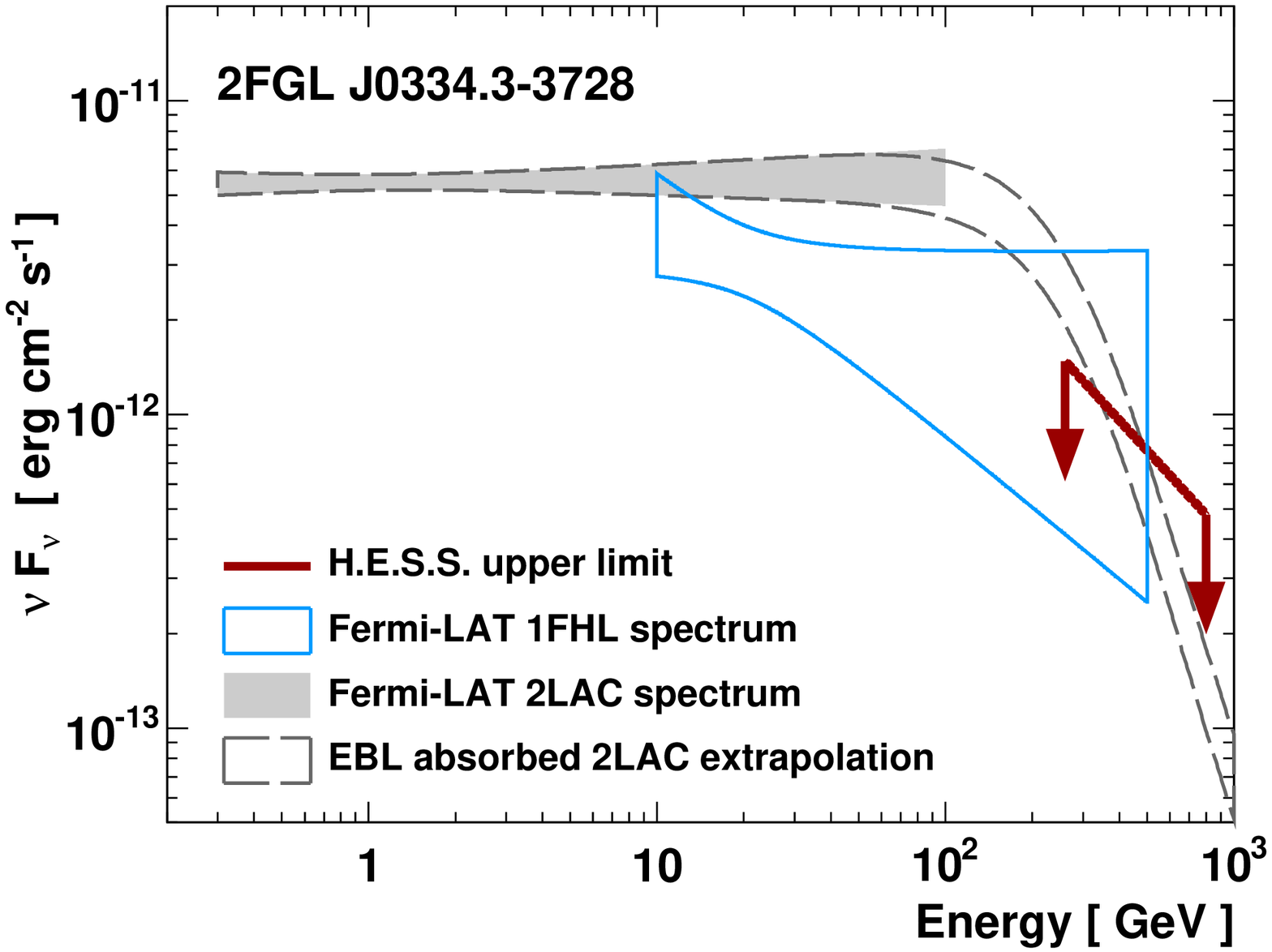}
\caption{{\it Left:} HE $\gamma$-ray spectrum and VHE upper limit on the emission of BZB~J0543-5532 as measured by \fermi\ and \hess\ The EBL-absorbed HE extrapolation based on the 2LAC is shown with a dashed line. {\it Right:} HE $\gamma$-ray spectrum and VHE upper limit on the emission of 2FGL~J0334.3-3728. For this object, a fiducial redshift of $0.3$ is assumed in the extrapolation.}
\label{FIG:UL_SPEC}
\end{figure*}

When taking into account the EBL absorption, all but two of the HE extrapolations lie below the \hess\ upper limits, indicating that no intrinsic curvature is required to explain the observed spectral break.  BZB~J0543-5532 is an exception, with a VHE upper limit a factor of two lower than the EBL-absorbed extrapolation. A straight power-law extrapolation of the intrinsic emission is therefore rejected, suggesting an intrinsic break in the photon spectrum. This curvature is also suggested by the marginal agreement between the \hess\ upper-limit and the high-energy end of the spectrum from the \fermi\ Catalog of Sources Above 10 GeV \citep[1FHL,][]{2013arXiv1306.6772T}. Similar conclusions can be drawn for 2FGL~J0334.3-3728, though with smaller statistics from the 1FHL and under the assumption that the object is nearby ($z<0.3$). Tighter constraints on the distance of this source and an increased coverage with \fermi\ and \hess\ will allow for more definite conclusions on the intrinsic emission of the source.

With the launch of \fermi, the AGN observation strategy at VHE has partly shifted from a target selection based on radio and X-ray fluxes towards a selection based on extrapolations of HE spectra. It should be noted nonetheless that, based only on the latter criterion, a fourth of the sources listed in Table~\ref{Table:Detection} would not have been discovered. High-frequency-peaked BL Lac objects such as PKS~0548-322, SHBL~J001355.9-185406, 1ES~1312-423, 1ES~0229+200, and 1ES~0347-121 are indeed not listed in the 2LAC because of a hard but faint HE emission. 

\begin{figure}[h]
\centering
\includegraphics[width=0.8\linewidth]{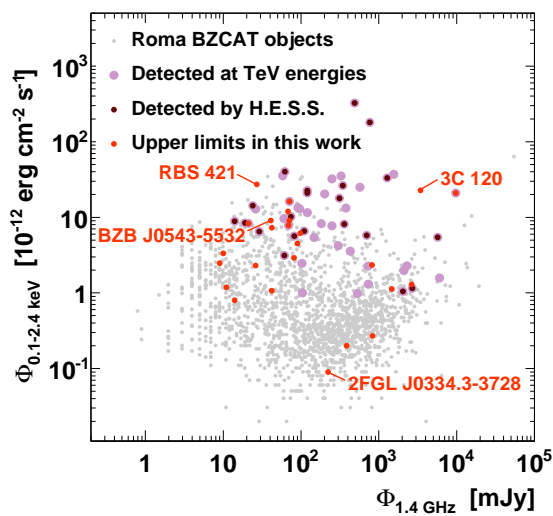} 
\caption{X-ray flux in the $\unit[0.1-2.4]{keV}$ band vs radio flux at $\unit[1.4]{GHz}$ for objects listed in the Roma BZCAT Catalog. 50 of the 56 AGN detected at VHE (as of the end 2013) are listed in the BZCAT with detected X-ray (ROSAT) and radio (NVSS/SUMSS) emission. 25 of the 47 objects studied in this paper are shown, with a selection biased toward X-ray-bright BL Lac objects. {\it Based on ROMA BZCAT and TeVCat.}}
\label{FIG:XrayRadio}
\end{figure}

Broadband multiwavelength strategies prove to be of critical importance in such cases. As discussed in \cite{2002AA...384...56C} and illustrated in Fig.~\ref{FIG:XrayRadio}, bright TeV BL Lac objects tend to have bright X-ray and radio counterparts. The latter criterion is not sufficient, however, as it tends to discard FSRQs with their low X-ray fluxes (low-energy component peaking in the optical-infrared band) and as X-ray bright objects, such as RBS~421 or 3C~120 studied in this paper, do not necessarily show bright TeV counterparts. Good HE-based candidates do not necessarily cluster in the upper-right corner of Fig.~\ref{FIG:XrayRadio} either, as shown by the relatively low radio flux of BZB~J0543-5532 and the low X-ray flux of 2FGL~J0334.3-3728.

The extension of the population of AGN detected at VHE and the discovery of new types of sources will be a primary task of the future Cherenkov Telescope Array, CTA \citep{2013APh....43..215S,2013APh....43..103R}. A best-suited target selection will account both for multiwavelength information from radio, X-ray, optical, and HE instruments, and for the charting effort led by previous and current-generation VHE instruments.

\section{Conclusion} \label{Ccl}

A large sample of AGN has been observed with \hess\ since 2002, resulting in the discovery of more than a third of the known extragalactic VHE emitters. Observations of 47 targets without significant excess were selected and upper limits on their integral fluxes were computed. For 41 of these objects, the upper limits derived in this paper are the strongest to date. 

No significant flaring event was detected during the $\sim\unit[400]{h}$ of observation of the 47 targets. The straight extrapolation of the HE emission is challenged by the VHE upper limit for a dozen objects. For all but two of them, this spectral curvature can be accounted for by the interaction of $\gamma$~rays with the EBL.

Active galactic nuclei observations, which are crucial both for the understanding of the EBL and of the objects themselves, will remain a primary goal of \hess\ during its second phase, \hess~II, where observations at lower energies will increase the number of detected sources and the maximum redshift accessible by Cherenkov telescopes. Extensive campaigns probing the sky down to fractions of percent of the Crab Nebula flux remain a major task of current VHE telescopes. This tremendous effort is paving the way for targeted AGN observations with CTA.

\begin{acknowledgements}
The support of the Namibian authorities and of the University of Namibia
in facilitating the construction and operation of H.E.S.S. is gratefully
acknowledged, as is the support by the German Ministry for Education and
Research (BMBF), the Max Planck Society, the German Research Foundation (DFG), 
the French Ministry for Research,
the CNRS-IN2P3 and the Astroparticle Interdisciplinary Programme of the
CNRS, the U.K. Science and Technology Facilities Council (STFC),
the IPNP of the Charles University, the Czech Science Foundation, the Polish 
Ministry of Science and  Higher Education, the South African Department of
Science and Technology and National Research Foundation, and by the
University of Namibia. We appreciate the excellent work of the technical
support staff in Berlin, Durham, Hamburg, Heidelberg, Palaiseau, Paris,
Saclay, and in Namibia in the construction and operation of the
equipment.

This research has made use of the SIMBAD database, operated at CDS, Strasbourg, France.
\end{acknowledgements}

\bibliography{UL_HESS.bbl}
\bibliographystyle{aa}

\end{document}